\documentclass[twocolumn]{aastex63}
\hypersetup{linkcolor=red,citecolor=blue,filecolor=cyan,urlcolor=black}

\usepackage{footnote}

\newcommand{\kms}{\ifmmode\,{\rm km}\,{\rm s}^{-1}\else km$\,$s$^{-1}$\fi}

\newcommand{\magarc}{\ifmmode {{{{\rm mag}~{\rm arcsec}}^{-2}}}
             \else {{{mag}$~${arcsec}$^{-2}$}}
             \fi}
\newcommand{\hunit}{km~s$^{-1}$~Mpc$^{-1}$}
\newcommand{\Nunit}{cm$^{-3}$}

\newcommand{\Lya}{Ly$\alpha$}

\newcommand{\funit}{\mathrm{ergs}~\mathrm{s}^{-1}~\mathrm{cm}^{-2}}
\newcommand{\Msun}{\mathrm{M}_{\sun}}
\newcommand{\Msunyr}{\mathrm{M}_{\sun}~\mathrm{yr}^{-1}}

\newcommand{\Hii}{H~{\sc ii}}

\newcommand{\Heii}{He~{\sc ii}}
\newcommand{\Oii}{[O~{\sc ii}]}
\newcommand{\Oiii}{[O~{\sc iii}]}
\newcommand{\Oiiit}{[O~{\sc iii}]~$\lambda$4363}
\newcommand{\OiiiUV}{O~{\sc iii}]}
\newcommand{\Ciii}{C~{\sc iii}]}
\newcommand{\Civ}{C~{\sc iv}}
\newcommand{\Mgii}{Mg~{\sc ii}}

\newcommand{\Ha}{H$\alpha$}
\newcommand{\Hb}{H$\beta$}
\newcommand{\Hg}{H$\gamma$}
\newcommand{\Hd}{H$\delta$}

\newcommand{\Hp}{H$^{+}$}
\newcommand{\Op}{O$^{+}$}
\newcommand{\Opp}{O$^{++}$}
\newcommand{\Oppp}{O$^{3+}$}
\newcommand{\Cp}{C$^{+}$}
\newcommand{\Cpp}{C$^{++}$}
\newcommand{\Cppp}{C$^{3+}$}
\newcommand{\Spp}{S$^{++}$}
\newcommand{\OH}{$12 + \mathrm{\log{(O/H)}}$}
\newcommand{\OHpp}{$12 + \mathrm{\log{(O^{++}/H^{+})}}$}

\newcommand{\Te}{T$_{\mathrm e}$}
\newcommand{\Ne}{n$_{\mathrm e}$}

\newcommand{\hst}{\textit{HST}}
\newcommand{\jwst}{\textit{JWST}}

\newcommand{\cgs}{erg\,s$^{-1}$\,cm$^{-2}$}

\newcommand{\um}{$\mu$m}

\usepackage{graphicx}

\graphicspath{{./}{figures/}}

\shorttitle{Low C/O abundance in a $z=6.23$ galaxy}
\shortauthors{Jones et al.}

\begin{document}

\title{Early Results from GLASS-JWST. XXI: Rapid assembly of a galaxy at $z=6.23$ revealed by its C/O abundance}

\correspondingauthor{Tucker Jones}
\email{tdjones@ucdavis.edu}

\author[0000-0001-5860-3419]{Tucker Jones}
\affiliation{Department of Physics and Astronomy, University of California Davis, 1 Shields Avenue, Davis, CA 95616, USA}

\author[0000-0003-4792-9119]{Ryan Sanders}
\altaffiliation{NHFP Hubble Fellow}
\affiliation{Department of Physics and Astronomy, University of California Davis, 1 Shields Avenue, Davis, CA 95616, USA}

\author[0000-0003-4520-5395]{Yuguang Chen}
\affiliation{Department of Physics and Astronomy, University of California Davis, 1 Shields Avenue, Davis, CA 95616, USA}

\author[0000-0002-9373-3865]{Xin Wang}
\affiliation{School of Astronomy and Space Science, University of Chinese Academy of Sciences (UCAS), Beijing 100049, China}
\affiliation{National Astronomical Observatories, Chinese Academy of Sciences, Beijing 100101, China}
\affiliation{Institute for Frontiers in Astronomy and Astrophysics, Beijing Normal University, Beijing 102206, China}

\author[0000-0002-8512-1404]{Takahiro Morishita}
\affiliation{Infrared Processing and Analysis Center, Caltech, 1200 E. California Blvd., Pasadena, CA 91125, USA}

\author[0000-0002-4140-1367]{Guido Roberts-Borsani}
\affiliation{Department of Physics and Astronomy, University of California, Los Angeles, 430 Portola Plaza, Los Angeles, CA 90095, USA}

\author[0000-0002-8460-0390]{Tommaso Treu}
\affiliation{Department of Physics and Astronomy, University of California, Los Angeles, 430 Portola Plaza, Los Angeles, CA 90095, USA}

\author{Alan Dressler}
\affiliation{The Observatories, The Carnegie Institution for Science, 813 Santa Barbara St., Pasadena, CA 91101, USA}

\author[0000-0001-6870-8900]{Emiliano Merlin}
\affiliation{INAF Osservatorio Astronomico di Roma, Via Frascati 33, 00078 Monteporzio Catone, Rome, Italy}

\author[0000-0002-7409-8114]{Diego Paris}
\affiliation{INAF Osservatorio Astronomico di Roma, Via Frascati 33, 00078 Monteporzio Catone, Rome, Italy}

\author[0000-0002-9334-8705]{Paola Santini}
\affiliation{INAF Osservatorio Astronomico di Roma, Via Frascati 33, 00078 Monteporzio Catone, Rome, Italy}

\author[0000-0003-1383-9414]{Pietro Bergamini}
\affiliation{Dipartimento di Fisica, Università degli Studi di Milano, Via Celoria 16, I-20133 Milano, Italy}
\affiliation{INAF - OAS, Osservatorio di Astrofisica e Scienza dello Spazio di Bologna, via Gobetti 93/3, I-40129 Bologna, Italy}

\author[0000-0002-6586-4446]{A. Henry}
\affiliation{Space Telescope Science Institute, 3700 San Martin Drive, Baltimore MD, 21218} 
\affiliation{Center for Astrophysical Sciences, Department of Physics and Astronomy, Johns Hopkins University, Baltimore, MD, 21218} 

\author{Erin Huntzinger}
\affiliation{Department of Physics and Astronomy, University of California Davis, 1 Shields Avenue, Davis, CA 95616, USA}





\author[0000-0003-2804-0648 ]{Themiya Nanayakkara}
\affiliation{Centre for Astrophysics and Supercomputing, Swinburne University of Technology, PO Box 218, Hawthorn, VIC 3122, Australia}



\author[0000-0003-4109-304X]{Kristan Boyett}
\affiliation{School of Physics, University of Melbourne, Parkville 3010, VIC, Australia}
\affiliation{ARC Centre of Excellence for All Sky Astrophysics in 3 Dimensions (ASTRO 3D), Australia}

 \author[0000-0001-5984-0395]{Marusa Bradac}
 \affiliation{University of Ljubljana, Department of Mathematics and Physics, Jadranska ulica 19, SI-1000 Ljubljana, Slovenia}
 \affiliation{Department of Physics and Astronomy, University of California Davis, 1 Shields Avenue, Davis, CA 95616, USA}

 \author[0000-0003-2680-005X]{Gabriel Brammer}
 \affiliation{Cosmic Dawn Center (DAWN), Denmark}
 \affiliation{Niels Bohr Institute, University of Copenhagen, Jagtvej 128, DK-2200 Copenhagen N, Denmark}

\author[0000-0003-2536-1614]{Antonello Calabr\'o}
\affiliation{INAF Osservatorio Astronomico di Roma, Via Frascati 33, 00078 Monteporzio Catone, Rome, Italy}



\author[0000-0002-3254-9044]{Karl Glazebrook}
\affiliation{Centre for Astrophysics and Supercomputing, Swinburne University of Technology, PO Box 218, Hawthorn, VIC 3122, Australia}

\author[0000-0002-3247-5321]{Kathryn~Grasha}
\altaffiliation{ARC DECRA Fellow}
\affiliation{Research School of Astronomy and Astrophysics, Australian National University, Canberra, ACT 2611, Australia}   
\affiliation{ARC Centre of Excellence for All Sky Astrophysics in 3 Dimensions (ASTRO 3D), Australia}   





 \author[0000-0002-9572-7813]{Sara Mascia}
 \affiliation{INAF Osservatorio Astronomico di Roma, Via Frascati 33, 00078 Monteporzio Catone, Rome, Italy}





 \author[0000-0001-8940-6768 ]{Laura Pentericci}
 \affiliation{INAF Osservatorio Astronomico di Roma, Via Frascati 33, 00078 Monteporzio Catone, Rome, Italy}





\author[0000-0001-9391-305X]{Michele Trenti} \affiliation{School of Physics, University of Melbourne, Parkville 3010, VIC, Australia}
\affiliation{ARC Centre of Excellence for All Sky Astrophysics in 3 Dimensions (ASTRO 3D), Australia}


\author[0000-0003-0980-1499]{Benedetta Vulcani}
\affiliation{INAF- Osservatorio astronomico di Padova, Vicolo Osservatorio 5, I-35122 Padova, Italy}

\begin{abstract}

The abundance of carbon relative to oxygen (C/O) is a promising probe of star formation history in the early universe, as the ratio changes with time due to production of these elements by different nucleosynthesis pathways. 
We present a measurement of $\log{\mathrm{(C/O)}} = -1.01\pm0.12$ (stat) $\pm0.15$ (sys) in a $z=6.23$ galaxy observed as part of the GLASS-JWST Early Release Science Program. Notably, we achieve good precision thanks to the detection of the rest-frame ultraviolet \OiiiUV, \Ciii, and \Civ\ emission lines delivered by \jwst/NIRSpec. 
The C/O abundance is $\sim$0.8 dex lower than the solar value and is consistent with the expected yield from core-collapse supernovae, 
indicating that longer-lived intermediate mass stars have not fully contributed to carbon enrichment.
This in turn implies rapid buildup of a young stellar population with age $\lesssim100$~Myr in a galaxy seen $\sim$900 million years after the Big Bang. 
Our chemical abundance analysis is consistent with spectral energy distribution modeling of \jwst/NIRCam photometric data, which indicates a current stellar mass $\log\,\mathrm{M}_* /\Msun = 8.4^{+0.4}_{-0.2}$ and specific star formation rate sSFR~$\simeq 20$~Gyr$^{-1}$.
These results showcase the value of chemical abundances and C/O in particular to study the earliest stages of galaxy assembly. 

\end{abstract}


\section{Introduction}
\label{sec:introduction}

The abundance of heavy elements is a fundamental property of galaxies that traces their growth and star formation, since metals originate from nucleosynthesis in the stellar evolution process \citep[e.g.,][]{maiolino2019,matteucci2012}.
The abundance pattern of metals with different nucleosynthetic origins and enrichment timescales provides a particularly powerful tool for constraining the formation history of galaxies. 
Since different elements can originate from stars of different masses, the timescales on which the interstellar medium (ISM) is enriched with these elements will differ according to the variation of stellar lifetime with mass. 
In the simple ``closed-box'' chemical evolution model,
oxygen and other $\alpha$ elements are predominantly produced in massive stars (M~$\gtrsim8~\Msun$) and returned to the ISM on short timescales by core-collapse supernovae (SNe; $\sim$10~Myr).
While carbon is also produced in massive stars, another important pathway is via intermediate-mass (M~$\sim1-4~\Msun$) asymptotic giant branch (AGB) stars with lifetimes of $\sim100~\mbox{Myr}-10~\mbox{Gyr}$ \citep[e.g.,][]{kobayashi2011,kobayashi2020}.
Consequently, galaxies with a formation timescale of $\lesssim100$~Myr will have C/O abundance approximately equivalent to the yield from core-collapse SNe alone, while C/O increases at ages $\gtrsim100$~Myr. 
The gas-phase abundance ratio C/O can thus indicate whether the stellar population in a galaxy is dominated by stars younger or older than $\sim$100~Myr.

C/O abundance is a promising tracer of the earliest phases of galaxy formation both because of its variation on relatively short timescales ($\sim$100~Myr) and because it can be derived from ratios of rest-frame ultraviolet (UV) emission lines of C (\Ciii$\lambda\lambda$1907,1909, \Civ$\lambda\lambda$1548,1550) and O (\OiiiUV$\lambda\lambda$1661,1666). These are typically the brightest UV nebular emission lines \citep[e.g.,][]{berg2022,byler2018} and are accessible with \jwst/NIRSpec out to extremely high redshifts ($z\sim4-30$). These lines have a further advantage that their ratios are relatively insensitive to dust reddening due to their close proximity in wavelength. 
Measurements of \Ciii, \Civ, and \OiiiUV\ at $z>6$ have shown that these high-ionization lines are strong in $z>6$ sources, with equivalent widths significantly larger than all but the most extreme galaxies at $z\sim0$ \citep[e.g.,][]{stark15a,stark15b,stark17,mainali2017,senchyna17,hutchison2019}.
The C/O abundance ratio is thus a premier tool for inferring the formation timescales of galaxies in the epoch of reionization during the first billion years of cosmic history.

In the local universe, a relation has been found between C/O and O/H in which C/O plateaus to a low-metallicity floor which averages log(C/O$)\sim-0.7$ at 12+log(O/H$)<8.0$, the primary nucleosynthesis regime, while C/O increases with O/H at higher metallicity \citep[e.g.,][]{henry2000,esteban2014,berg16,berg2019,tsc17}.
Using rest-UV spectroscopy of nearby dwarf galaxies, \citet{berg2019} showed that this relation displays a large intrinsic scatter of $\sim0.2$~dex in C/O at fixed O/H.
This scatter has been shown to be a function of the star formation history (for the enrichment timescale reasons outlined above) and the preferential removal of O relative to C by SNe-driven outflows, where galaxies with shorter formation timescales and little preferential O removal have lower C/O \citep{berg2019,yin2011}.
Intermediate-redshift galaxies at $z\sim2-3$ fall on the low-metallicity plateau with a mean value and scatter similar to the $z\sim0$ sample \citep[e.g.,][]{berg2018,berg2019}.
\citet{arellano22} recently reported the first C/O determination at $z>6$ enabled by \jwst\ spectroscopy.
These authors found log(C/O$)=-0.83\pm0.38$ for a $z=8.495$ galaxy, consistent with the local low-metallicity plateau. However, the interpretation of this value is clouded by its low precision due to a marginal detection of \Ciii\ (2.4$\sigma$) and no detection of rest-UV \OiiiUV, relying instead on the ratio relative to rest-optical \Oiii\ lines that is highly sensitive to dust reddening. This early result nonetheless provides a precise O/H abundance and shows the promise of \jwst\ spectroscopy for abundance patterns at extremely high redshifts. 

In this work, we present the first high-precision measurement of C/O for a galaxy at $z=6.23$ enabled by \jwst/NIRSpec measurements of the rest-UV \Ciii, \Civ, and \OiiiUV\ lines from the GLASS-JWST ERS program \citep{treu22}. This target (source ID 150008 in the GLASS NIRSpec target catalog and ID 2649 in the Stage 1 photometric catalog described by \citealt{paris2023}; RA, Dec = 3.6025240, $-$30.4192187 degrees) was originally included in the NIRSpec observations as a candidate $z>5$ galaxy based on photometry indicating a Lyman break. 
We visually inspected the GLASS-JWST spectra of known candidate $z>5$ sources for suitability of C/O abundance measurements and selected this as the best example based on clear detection of the necessary rest-UV lines. For many otherwise promising galaxies, either the \Ciii\ or \OiiiUV\ lines are not covered due to their position on the slitmask (including the $z\sim8$ protocluster members described in \citealt{morishita2022}). Our current work represents a high-redshift case study which also serves to illustrate the value and feasibility of a future enlarged sample. 

This paper is organized as follows.
In Section~\ref{sec:data}, we describe the observations, data reduction, and photometric and spectroscopic measurements. In Section~\ref{sec:analysis} we derive the physical properties of this target, including the electron temperature (Sec.~\ref{sec:Te}), ionic and total abundance ratios (Sec.~\ref{sec:abundances}), and stellar population properties (Sec.~\ref{sec:SED}).
We discuss the results and present our conclusions in Section~\ref{sec:conclusions}. 

Throughout this work we adopt the concordance $\Lambda$CDM cosmology with $\mathrm{H}_{0}=70$~\hunit, $\Omega_{m}=0.3$, and $\Omega_{\Lambda}=0.7$. 
We use atomic data from \cite{tz17} for \Opp\ collision strengths, \cite{fft04} for \Opp\ transition probabilities, \cite{ber85} for \Cpp\ collision strengths, \cite{ak04} for \Cppp\ collision strengths, and \cite{wfd96} for \Cpp\ and \Cppp\ transition probabilities. We adopt the solar abundance pattern of \cite{asplund21}. 


\section{Observations}
\label{sec:data}

\subsection{Photometry}
\label{sec:photometry}

We use 7-band \jwst/NIRCam photometry from the UNCOVER program (JWST-GO-2561; \citealt{bezanson2022}) to constrain the stellar population and star formation history. The data reduction and measurement methods are as described in \cite{merlin22} and \cite{paris2023}; here we give a brief summary. 
The mosaics in all bands are obtained using a customised version of the STScI pipeline for \jwst\ (\texttt{CRDS\_VER} 11.16.14, \texttt{CAL\_VER} 1.8.2), with tailored modules to accurately perform the astrometric alignment and to remove defects such as snowballs, wisps, and claws \citep[see][]{rigby2022}. Sources are detected on the F444W image using \textsc{SExtractor} \citep{bertin96}. To extract the multiband photometry, images from each photometric filter are first convolved to match the F444W filter's point spread function. Colors are then measured within an 0\farcs28 circular aperture (twice the FWHM of the F444W image) using \textsc{a-phot} \citep{merlin2019}. The total F444W flux is calculated within a Kron elliptical aperture, with fluxes in other bands given by the 0\farcs28-aperture color scaled to the total flux. The results are given in Table~\ref{tab:fluxes}. 

At the redshift $z=6.23$ of our target, F356W and F444W broad-band fluxes include strong nebular emission lines (\Hb+\Oiii\ and \Ha, respectively). This is clearly apparent in the photometry (Table~\ref{tab:fluxes}), with flux density in these bands elevated by a factor $\sim$1.5. The difference of $\sim $0.07~$\mu$Jy in F356W compared to adjacent filters suggests approximately $1.3\times10^{-17}~\funit$ of emission line flux contribution, in reasonable agreement (within $\sim$10\%) with the measured fluxes of \Oiii\ and \Hb\ (Section~\ref{sec:spectroscopy}). 
The F410M filter is relatively unaffected by nebular emission and provides a reliable measurement of stellar continuum at rest-frame $\sim$5700~\AA. Overall the NIRCam photometry provides good sampling of the rest-frame UV through optical continuum ($\simeq$1400--7000~\AA) including the Balmer and 4000~\AA\ breaks. 
We additionally include fluxes from the Hubble Space Telescope (\hst) via the Frontier Fields program (\citealt{lotz17,merlin2016,castellano16}; data are available in MAST: \dataset[10.17909/t9-4xvp-7s45]{http://dx.doi.org/10.17909/t9-4xvp-7s45}) which sample across the Lyman break at $z=6.23$. However, the \hst\ photometry has little effect on results in this paper.

\begin{table}
\centering
\begin{tabular}{lcc}
\hline
{Transition} & {Flux} & {FWHM} \\
  & ($10^{-18}$~\cgs) & (\AA) \\
\hline
\hline
\Oiii\ $\lambda$5007\tablenotemark{a} &  9.53 $\pm$  0.08  &   16.4 $\pm$   0.2  \\
\Hb\tablenotemark{a}                  &  1.71 $\pm$  0.07  &  \\
\Hd\                 &  0.38 $\pm$  0.05  &   12.2 $\pm$   1.9    \\
\Ciii  $\lambda$1909\tablenotemark{b} &  0.20 $\pm$  0.05  \\
\Ciii  $\lambda$1907\tablenotemark{b} &  0.36 $\pm$  0.06  &    5.3 $\pm$   0.9  \\
\OiiiUV~$\lambda$1666\tablenotemark{c} &  0.53 $\pm$  0.08  &    8.3 $\pm$   1.4  \\
\Heii\ $\lambda$1640\tablenotemark{c} &  0.09 $\pm$  0.07  \\
\Civ   $\lambda$1551\tablenotemark{d} &  0.22 $\pm$  0.06  \\
\Civ   $\lambda$1549\tablenotemark{d} &  0.23 $\pm$  0.06  &    4.5 $\pm$   1.1  \\
\hline
\\
Filter  &  $f_{\nu}$ ($\mu$Jy)  \\
\hline
\hline
%
F435W  & 0.001 $\pm$ 0.002 \\
F606W  & 0.001 $\pm$ 0.003 \\
F814W  & 0.016 $\pm$ 0.002 \\
F105W  & 0.113 $\pm$ 0.027 \\
F125W  & 0.110 $\pm$ 0.024 \\
F160W  & 0.091 $\pm$ 0.031 \\
\hline
F115W  & 0.106 $\pm$ 0.005 \\
F150W  & 0.109 $\pm$ 0.004 \\
F200W  & 0.118 $\pm$ 0.004 \\
F277W  & 0.109 $\pm$ 0.003 \\
F356W  & 0.179 $\pm$ 0.003 \\
F410M  & 0.113 $\pm$ 0.005 \\
F444W  & 0.172 $\pm$ 0.008 \\
\hline

\end{tabular}
\tablenotetext{a}{Joint fit of \Oiii\ $\lambda\lambda$4959,5007 and \Hb.}
\tablenotetext{b}{Joint fit of \Ciii\ $\lambda\lambda$1907,1909.}
\tablenotetext{c}{Joint fit of \OiiiUV\ $\lambda\lambda$1661,1666 and \Heii\ $\lambda$1640.}
\tablenotetext{d}{Joint fit of \Civ\ $\lambda\lambda$1549,1551.}
\caption{Emission line fluxes and photometry. Line widths are given as the FWHM from Gaussian fits with no correction for the instrument line spread function. In cases where multiple lines are fit jointly (with the same width and redshift), the FWHM is reported only for the strongest line.
Photometric flux densities $f_{\nu}$ are measured following the methods described in \cite{merlin22} and \cite{paris2023}. The top measurements in filters F435W through F160W are from \hst, while F115W and below are from \jwst/NIRCam.}
\label{tab:fluxes}
\end{table}

\subsection{NIRSpec spectroscopy and line fluxes}
\label{sec:spectroscopy}

\begin{figure*}
\center
 \includegraphics[width=2.2in]{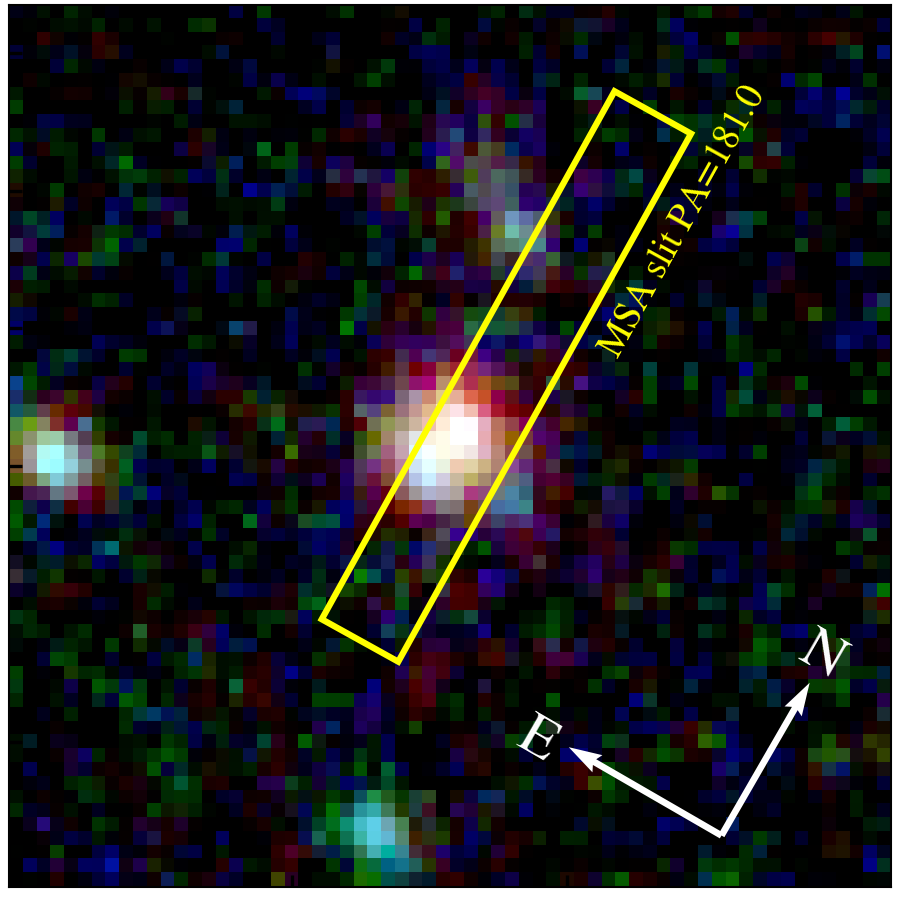}
 \includegraphics[width=2.3in]{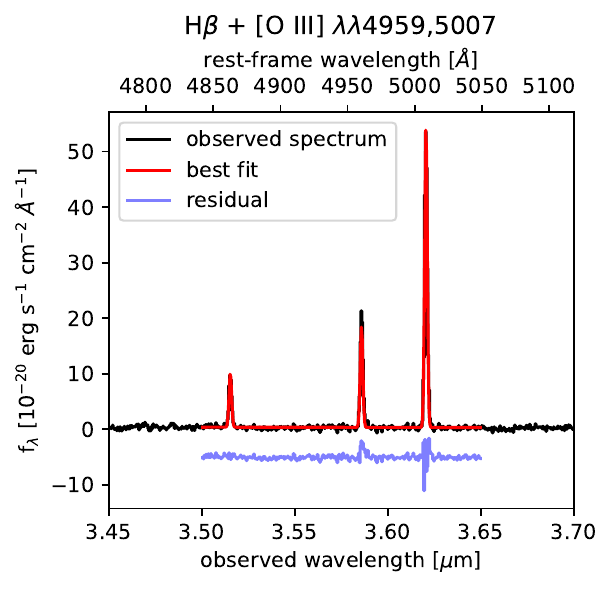}
 \includegraphics[width=2.3in]{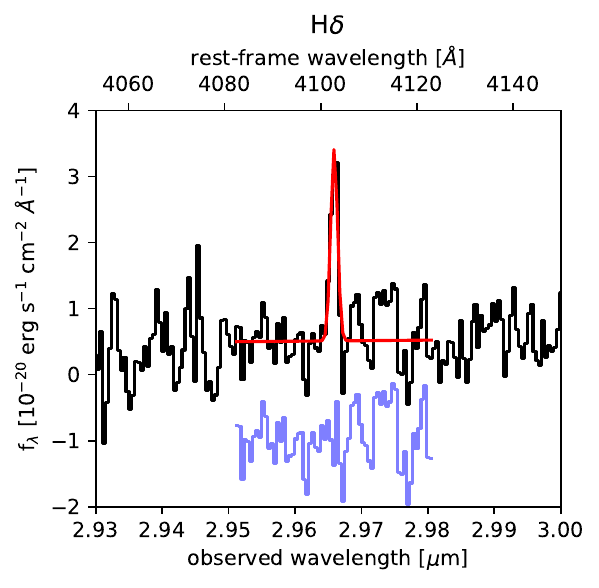}
 \\
 \includegraphics[width=2.3in]{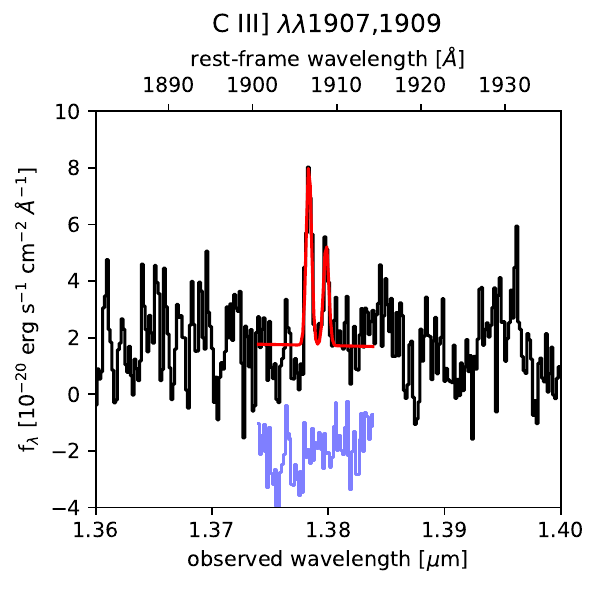}
 \includegraphics[width=2.3in]{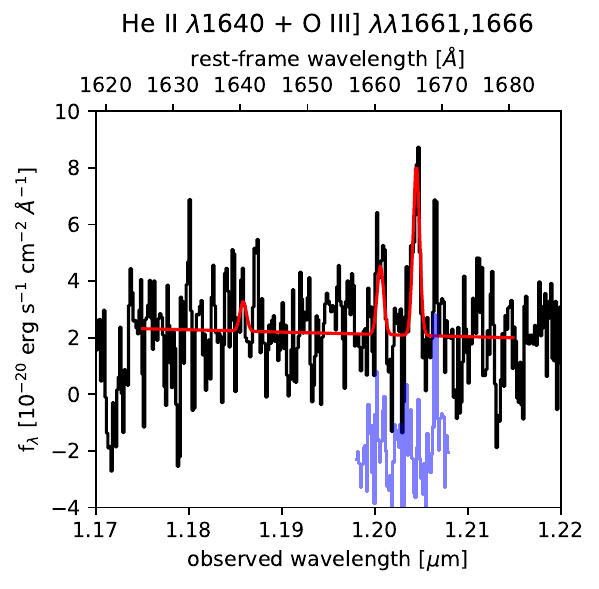}
 \includegraphics[width=2.3in]{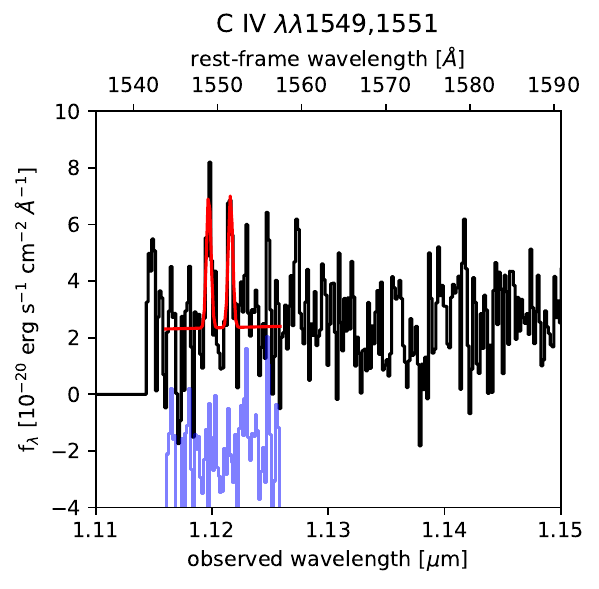}
 \caption{Overview of the spectroscopic data. The top left panel shows \jwst/NIRCam imaging (F115W/F200W/F444W = blue/green/red) in a 2\arcsec$\times$2\arcsec field centered on the target galaxy, with the position of the NIRSpec slitlet indicated by the rectangle. The other panels show regions of the spectra around the main emission lines used in this work with best-fit profiles shown in red (see Section~\ref{sec:spectroscopy} for details). Residuals around each line are shown in blue, offset vertically for clarity. All key lines used in this work are clearly detected, with $\gtrsim7\sigma$ significance for the rest-UV \OiiiUV\ and \Ciii\ doublets and $5.2\sigma$ for the \Civ\ doublet. Residuals are generally consistent with the noise level in the spectra, except for \Oiii~$\lambda\lambda$4959,5007 which is discussed in Section~\ref{sec:spectroscopy}.}
 \label{fig:line_fits}
\end{figure*}

We obtained moderate resolution ($R=\lambda/\Delta\lambda\simeq2700$) spectroscopy covering $\lambda_{\rm obs}\simeq1.0$--5.3~\um\ with \jwst/NIRSpec in multi-object spectroscopy (MOS) mode as part of the GLASS-JWST survey (ERS 1324, PI Treu; \citealt{treu22}; see also \citealt{morishita2022} for details of the NIRSpec observations). The slitlet position for our target is shown in Figure~\ref{fig:line_fits}. 
We reduced the NIRSpec spectra using a combination of the default STScI \jwst\ calibration pipeline and the \textsc{msaexp} software\footnote{\url{https://github.com/gbrammer/msaexp}}. First, count-rate maps are produced from the uncalibrated data using \textsc{calwebb\_detector1} with the most recent available reference files ({\sc jwst\_1014.pmap}). Then \textsc{msaexp} conducts additional preprocessing steps to remove the 1/f noise and ``snowball'' features in the rate images, and calls the level-2 \textsc{calwebb\_spec2} reduction scripts to extract 2D spectra from individual exposures, after WCS registration, slit path-loss correction, flat-fielding, wavelength and flux calibrations.
Subsequently, \textsc{msaexp} performs an optimal 1D spectral extraction based on the \cite{Horne.1986} algorithm, utilizing the target light profile along the cross-dispersion direction for the optimal extraction aperture.
Finally, the 1D spectra extracted from multiple exposures at various dither positions and visits are combined via median stacking with outlier rejections.
Our target has well-detected continuum traces in individual exposures, making it feasible to extract and combine the 1D spectra as opposed to first combining the 2D spectra. This method is advantageous for bright objects with sub-pixel dithering, since it enables oversampling the line spread function of our NIRSpec observations in order to improve the sampling of the emission line profiles.

Given the location of our target on the slitmask, the observed wavelength coverage is approximately 1.0--1.6~\um\ with F100LP/G140H, 1.7--2.65~\um\ with F170LP/G235H, and 2.9--4.5~\um\ with F290LP/G395H. There is also a $\sim$0.1~\um\ detector gap near the short-wavelength end of each range.
This range corresponds to $\sim$1360--6220~\AA\ in the rest frame.
The spectra include several key rest-frame UV lines used in this analysis (\Ciii~$\lambda\lambda$1907,1909, \OiiiUV~$\lambda\lambda$1661,1666, \Civ~$\lambda\lambda$1549,1551) as well as prominent rest-frame optical lines (\Hd, \Hb, \Oiii~$\lambda\lambda$4959,5007), shown in Figure~\ref{fig:line_fits}. 
\Hg\ and \Oiiit\ are not covered due to the detector gap, while \Ha\ and \Oii~$\lambda\lambda$3727,3729 fall redward of the detector area for the slitmask position. Although these missing lines would be useful, the available spectral coverage is suitable for our goal of measuring the C/O abundance. 

To account for uncertainties in flux calibration, slit loss, or other factors, we scale the observed spectra to match the photometric flux densities (Section~\ref{sec:photometry}). The median spectroscopic continuum value is measured within 0.05~\um\ of the central wavelength for F150W (for the G140H spectrum), 0.07~\um\ for F200W (G235H spectrum), and 0.10~\um\ for F410M (G395 spectrum). These filters and spectral ranges are chosen to sample representative parts of each spectral tuning, avoiding strong emission lines. The resulting signal-to-noise is $\gtrsim20$ in the median continuum values. 
The spectra from each grating are then scaled such that these median flux densities match the photometric measurements (Table~\ref{tab:fluxes}). 
Notably the main result of C/O abundance derived in this paper is relatively unaffected by scaling effects such as flux calibration and slit losses, since the relevant rest-frame UV lines are close in wavelength and observed in the same grating (F100LP/G140H). 

Each emission line of interest is fit with a Gaussian profile along with a first-order polynomial to model the continuum, within a range $\Delta\lambda \simeq 0.1$~\um\ around the line centroid. The best-fit line fluxes and Gaussian full width at half maximum (FWHM) values are given in Table~\ref{tab:fluxes}, with line profiles shown in Figure~\ref{fig:line_fits}. 
We fit nearby lines jointly, such as the UV emission doublets and the optical \Oiii+\Hb, using the same redshift and Gaussian width for each line. 
We also impose the theoretically expected flux ratios \Oiii~$\lambda$5007/$\lambda$4959~$=2.98$ and \OiiiUV~$\lambda$1666/$\lambda$1661~$=2.49$ in these joint fits, and we report only the stronger line of each doublet. We do not impose constraints on the \Civ\ doublet flux ratio as it can be affected by resonant absorption and scattering, as well as P Cygni stellar features. 

In general we find that residuals from these fits are consistent with the noise level. The sole exception is the optical \Oiii\ doublet, for which the joint fit underestimates the flux of \Oiii~$\lambda$4959 by 15\% (and overestimates \Oiii~$\lambda$5007 by 2\%) compared to fitting the lines individually. Given this disagreement with the expected flux ratio, the true uncertainty in \Oiii~$\lambda$5007 flux may be as large as $\sim$10\% (cf. the $\sim$1\% statistical uncertainty reported in Table~\ref{tab:fluxes}). However, even if all lines in our analysis are subject to an additional 10\% uncertainty in flux, this would still be comparable or smaller than the statistical uncertainty in derived physical properties. 

\begin{figure}
\center
 \includegraphics[width=\columnwidth]{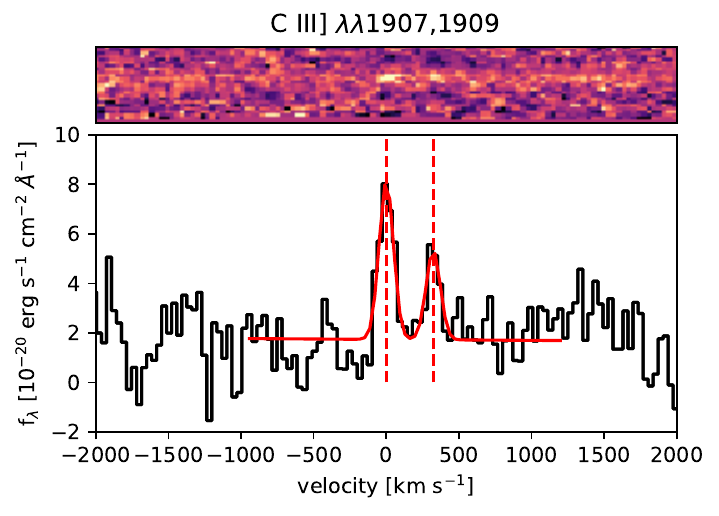}
 \includegraphics[width=\columnwidth]{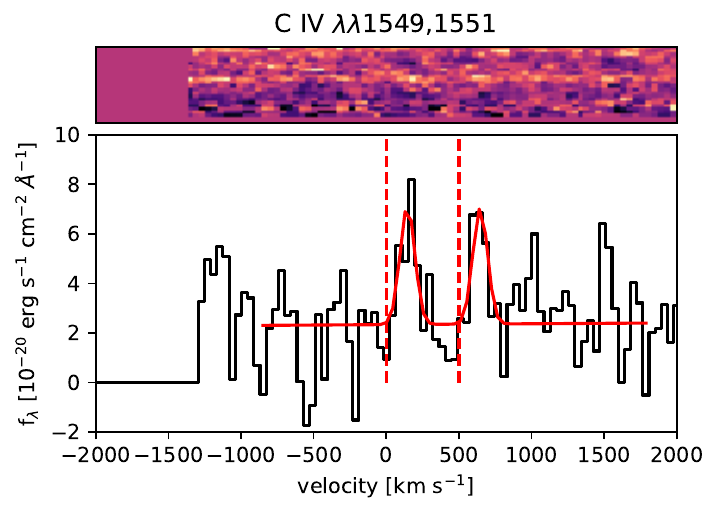}
 \caption{2D and 1D spectra of rest-frame UV emission lines in velocity space. Vertical dashed lines in the 1D panels show expected line centers for the \Ciii\ and \Civ\ doublets, relative to the best-fit redshift of optical \Oiii\ and \Hb. The velocity origin in each panel corresponds to the bluer line. Best-fit line profiles (Section~\ref{sec:spectroscopy}) are shown in red. The \Ciii\ redshift is consistent with the optical lines and rest-UV \OiiiUV, while \Civ\ emission is redshifted by $\sim$170~\kms. This \Civ\ velocity offset can be plausibly explained by resonant scattering in a galactic outflow.}
 \label{fig:line_velocities}
\end{figure}

The redshifts of each line fit provide a useful check of the wavelength calibration and uncertainty estimates. Excluding \Civ, the remaining four independent fits are all consistent within 1$\sigma$ of their weighted mean $\bar{z} = 6.22895\pm0.00007$ (with $\chi^2 = \sum \frac{(z - \bar{z})^2}{\sigma_z^2} = 1.6$ for 3 degrees of freedom). This indicates a reliable wavelength scale and good fitting results. 
We exclude \Civ\ emission from this mean because it shows a clear redward velocity shift of $171\pm14$~\kms\ relative to $\bar{z}$ (Figure~\ref{fig:line_velocities}). The \Civ\ emission appears to be real, exhibiting two lines (both at $>3\sigma$) at the expected doublet separation with $>5\sigma$ combined significance in our joint fit. 
We interpret this shift as arising from scattering in a galactic-scale outflow, which produces the commonly-observed redshifted emission in resonant lines (such as \Civ, \Mgii, and \Lya; e.g., \citealt{prochaska2011}). 
In this scenario we also expect \Civ\ absorption at velocities $v \lesssim 0$ from interstellar and outflowing gas along the line-of-sight. Consequently the \Civ\ emission flux can be affected by such absorption \citep[as described in detail by, e.g.,][]{senchyna2022}. 
However, the continuum signal-to-noise is such that we cannot obtain constraining measurements of interstellar absorption. Similarly we do not have strong constraints on the stellar P Cygni component, which can also affect the nebular emission line profile. 

We additionally consider the line widths as a test of the fit quality and as a dynamical mass estimator. The instrument resolution is FWHM~$\simeq14.7$~\AA\ in G395H (corresponding to \Hd, \Hb, and \Oiii) and 5.2~\AA\ in G140H (\Civ, \Ciii, \OiiiUV). The only fit with FWHM $>3\sigma$ above the instrument resolution is for \Hb\ and \Oiii, which gives an intrinsic velocity FWHM~$=60\pm4$~\kms\ corrected for the instrument resolution. All other fits agree within $2\sigma$ of this value, including \Hd\ and \Civ\ whose best-fit FWHM are smaller than the instrument resolution.
For \Heii~$\lambda$1640, the line width is fit jointly with \OiiiUV\ and the resulting fit should thus be interpreted as a nebular component (as opposed to, e.g., broad stellar emission). Regardless, \Heii\ emission is not detected ($\sim 1\sigma$). We note that excluding \Heii\ from the fit has negligible effect on the derived \OiiiUV\ flux or line width.
While we find an intrinsic velocity width FWHM~$\simeq60$~\kms, this may be an underestimate since the source appears to not fill the entire slit width (Figure~\ref{fig:line_fits}). The longer wavelength lines are likely more accurate as the coarser angular resolution will result in more uniform slit illumination. Our line width measurement is indeed based on the reddest lines available. The implied dynamical mass is $\sim 4\times10^{8}~\Msun$ within a radius of 1~kpc, with uncertainty of order a factor of 2 \citep[e.g.,][]{law2009}.

\section{Physical properties}
\label{sec:analysis}

In this section we present physical properties derived from the photometry and emission line measurements. Quantities such as the stellar mass and SFR must be corrected for the modest lensing magnification. We adopt a magnification factor $\mu = 2.7\pm0.1$ obtained from the lens model described in \cite{bergamini2022,bergamini2023}. This value and 1$\sigma$ confidence interval are determined from a Monte Carlo analysis, and account for the positional accuracy recovered in the updated model \citep{bergamini2023}.
The formal lensing uncertainties are small and we do not propagate them through the analysis, so that the derived properties can more easily be updated with predictions from other lens models. However, most of the relevant properties for this work (e.g., chemical abundances and stellar population age) are independent of lensing magnification.

\subsection{Nebular reddening, temperature and density}
\label{sec:Te}

The primary goal of this work is to determine the gas-phase abundances, particularly the C/O ratio. Here we use the direct \Te\ method which relies on nebular temperature (\Te) and density (\Ne) measurements. 

We first examine the nebular reddening in order to estimate \Te\ from the \Oiii\ emission lines. The Balmer flux ratio \Hd/\Hb\ is within 1.4$\sigma$ of the expected intrinsic value 0.264 (assuming Case B recombination with \Te~$=2\times10^4$~K and \Ne~$=200$~\Nunit). Table~\ref{tab:properties} lists the reddening E(B-V)~$=0.25^{+0.21}_{-0.18}$ based on the \cite{ccm89} attenuation law with $\mathrm{R_V}=4.05$ \citep[as found by][]{calzetti2000}. 
This does not include correction for underlying stellar absorption, which could plausibly reduce the E(B-V) value by $\sim$0.05 based on SED modeling results (Section~\ref{sec:SED}); this is well within the statistical uncertainty. 
The reddening is insensitive to temperature; an assumed range of \Te~$=(1.5-3)\times10^4$~K results in only $\pm0.01$ change to E(B-V). 
We additionally report the SFR derived from reddening-corrected \Hb\ luminosity using the \cite{kennicutt98} calibration corrected to a \cite{chabrier2003} initial mass function (IMF). 
While E(B-V) is relatively robust to the choice of attenuation law and $R_V$, we note that the total attenuation and UV-to-optical correction factor can vary by $\sim$50\%. We thus approach the reddening correction cautiously as the uncertainty is fairly large. The SED analyses described in Section~\ref{sec:SED} give best-fitting E(B-V)~$=0.1$--0.3 which, along with the SED-derived star formation rate, suggests that the true reddening is likely in the low end of our uncertainty range from \Hd/\Hb. 

We calculate electron temperature using \textsc{PyNeb} \citep{luridiana15} with the atomic data listed in Section~\ref{sec:introduction}. The result is \Te~$=24300^{+9600}_{-4700}$~K from the ratio of UV \OiiiUV\ to optical \Oiii\ emission, using the reddening correction described above (corresponding to a correction factor 2.1$\times$ in the UV-to-optical flux ratio) and assuming \Ne~=~200~\Nunit. (The result is insensitive to density; adopting an extreme $10^5$~\Nunit\ decreases \Te\ by only 2,000~K.) The uncertainty is entirely dominated by reddening correction, with only $\sim$1500~K uncertainty from emission line signal/noise. 
We can also place a firm lower limit \Te~$>$~18,000~K assuming no reddening. 
While this temperature is much higher than typical nearby galaxies and \Hii\ regions (which have \Te~$\sim$~10,000~K), high \Te\ is expected in metal-poor and highly star-forming galaxies \citep[e.g.,][]{perez-montero2021,berg2021}.
Indeeed, low-redshift extremely metal-poor galaxies (12+log(O/H$)<7.1$) have been found with \Te~$=$~21,000--25,000~K \citep[e.g.,][]{izotov2018,izotov2019,izotov2021}. 
While such high temperatures may be puzzling given the efficient cooling at \Te~$\gtrsim$~20,000 K,
comparable or higher \Te\ values have been reported in $z>7$ galaxies based on early \jwst\ observations \citep[e.g.,][]{curti2023,schaerer2022}, suggesting they may be common in the rapidly star forming population at this early epoch.

The \Ciii\ doublet ratio is sensitive to electron density. The measured 1907/1909 flux ratio of $1.78\pm0.47$ is compatible with a wide range of densities, though the best-fit measurement is formally unphysical (beyond the low-density limit). The 1$\sigma$ bounds indicate $n_e < 8200$~\Nunit\ assuming \Te~=~24,000~K. We reiterate that the change in derived \Te\ across this density range is insignificant for our analysis.

\begin{table}
\centering
\begin{tabular}{lc}
\hline
Property  &  Value  \\
\hline
\hline
RA   &  00:14:24.607  \\
Dec  &  $-$30:25:09.24  \\
$z$  &  $6.22895\pm0.00007$  \\
$\mu$  &  $2.7\pm0.1$  \\
$\log$~M$_*$ ($\Msun$)\tablenotemark{a}  &  $8.39 ^{+ 0.35 }_{- 0.19 }$ \\
SFR$_{\mathrm{SED}}$  ($\Msunyr$)\tablenotemark{a}  &  $5.1 ^{+ 6.2 }_{- 1.1 }$ \\
SFR$_{\mathrm{H\beta}}$  ($\Msunyr$)  &  $10^{+14} _{-5}$  \\
Age$_{\mathrm{par}}$ (Myr)\tablenotemark{a}  &  $126 ^{+ 375 }_{- 70 }$  \\
Age$_{\mathrm{non-par}}$ (Myr)\tablenotemark{b}  &  $99 ^{+ 132 }_{- 63 }$  \\
E(B-V)$_{\mathrm{gas}}$  &  $0.25^{+0.21}_{-0.18}$  \\
\Te\ (K)  &  $24300^{+9600}_{-4700}$  \\
\Ne\ (\Nunit)  &  $<8200$  \\
\hline
Abundances \\
\OHpp  &  $7.34^{+0.20}_{-0.22}$  \\
\OH\tablenotemark{c}  & $7.39^{+0.23}_{-0.20}$  \\
$\log($\Cpp/\Opp$)$  &  $-1.18 \pm 0.11$  \\
$[$\Cpp/\Opp$]$\tablenotemark{d}  &  $-0.95 \pm 0.11$  \\
$\log($\Cppp/\Cpp$)$  &  $-0.33 \pm 0.17$  \\
$\log{\left( \mathrm{\frac{C^{3+} + C^{++}}{O^{++}}} \right)}$  &   $-1.01 \pm 0.12$  \\
$\left[ \mathrm{\frac{C^{3+} + C^{++}}{O^{++}}} \right]$\tablenotemark{d}  &  $-0.78 \pm 0.12$  \\
$\log{\mathrm{(C/O)}}$  &  $-1.01 \pm 0.12~\mathrm{(stat)} \pm 0.15~\mathrm{(sys)}$  \\
\hline
\end{tabular}
\tablenotetext{a}{From the parametric SED fitting method of \cite{santini2022}.}
\tablenotetext{b}{From the non-parametric SED fitting method of \cite{morishita2019}.}
\tablenotetext{c}{Assuming O32 is uniformly distributed between $3.0-10$ and \Te(\Op)=\Te(\Opp).}
\tablenotetext{d}{Relative to solar $\log{\mathrm{(C/O)_{\odot}}} = -0.23$ \citep{asplund21}.}
\caption{Physical properties. The stellar mass (M$_*$) and SFR values are corrected for lensing magnification $\mu$, while other derived properties are not affected by lensing. For ion abundances such as \Cpp/\Opp\ we report the statistical uncertainties, excluding systematic uncertainty arising from possible differences in \Te\ associated with different ions (see discussion in Section~\ref{sec:abundances}).}
\label{tab:properties}
\end{table}

\subsection{Chemical abundances}
\label{sec:abundances}

We use the measured line fluxes and physical properties from Section~\ref{sec:Te} to calculate ionic abundances using \textsc{PyNeb}, and report the results in Table~\ref{tab:properties}. Our main interest is C/O. We measure \Cpp/\Opp\ ion abundance from the \Ciii~$\lambda\lambda$1907,1909 / \OiiiUV~$\lambda\lambda$1661,1666 flux ratio, which is relatively robust to various sources of uncertainty. 
The \Te\ uncertainty propagates to $\pm0.05$~dex in \Cpp/\Opp, with lower \Te\ corresponding to lower \Cpp/\Opp\ abundance. Flux measurement uncertainty (i.e., signal/noise) contributes $\pm0.06$ from each of the \OiiiUV\ and \Ciii\ doublets. Considering a range of density \Ne~$=1$--$10^3$~\Nunit\ corresponds to only $\pm0.03$ dex relative to our assumed 200~\Nunit\ 
(which is motivated by measurements at $z\gtrsim2$; e.g., \citealt{sanders2016}).
Although the high \Ne~$=8200$~\Nunit\ permitted by our data would increase the derived C/O by 0.08 dex, we also note that \Ciii-based densities are typically higher than found from the more widely-used optical diagnostics \citep[e.g.,][]{mingozzi2022}.
Reddening correction is somewhat difficult to assess, as a Milky Way-like attenuation law \citep[e.g., that adopted from][]{ccm89} indicates that \Ciii\ is more attenuated due to the strong 2175~\AA\ ``bump'' feature, while laws with weak or no bump \citep[e.g.,][]{calzetti2000,reddy15} instead have larger attenuation of \OiiiUV. 
While the bump feature tends to be weak in metal-poor galaxies, \cite{witstok2023} have recently reported a Milky Way-like bump strength in a low-metallicity $z\sim7$ galaxy. 
We thus conservatively adopt a relative reddening factor of $1.0\pm0.1$ (i.e., within 10\% of no reddening) for the \OiiiUV/\Ciii\ ratio, which encompasses the plausible range of attenuation laws given the E(B-V) value. This corresponds to uncertainty of only $\pm0.04$ dex, relatively small thanks to the small wavelength separation of the rest-UV emission lines. Collectively the total ion abundance uncertainty from sources discussed above is $\pm0.11$ dex, with the largest contribution from signal/noise of the rest-UV lines. 

Another source of uncertainty is the \textit{relative} temperature associated with \Ciii\ and \OiiiUV\ emission, which is not well established especially at the high \Te\ of our target. We specifically consider how abundance measurements are affected if \Ciii\ is associated with an intermediate-ionization zone characterized by \Te(\Spp), instead of \Te(\Opp). Extrapolating the \Te-\Te\ relation of \cite{garnett1992} gives an intermediate-ionization temperature lower by $\sim$2400~K, propagating to higher \Cpp/\Opp\ abundance (although Garnett suggests \Te(\Opp) is more appropriate for \Ciii\ emission). In contrast the \cite{croxall2016} and \cite{rogers2021} relations suggest the opposite with lower inferred \Cpp/\Opp. 
Given these relations, we consider a systematic uncertainty corresponding to $\pm 2400$~K difference in \Te(\Cpp) relative to \Te(\Opp). This propagates to $\pm0.15$~dex in \Cpp\ abundance relative to \Opp\ and \Cppp. The magnitude of this effect is therefore potentially comparable to the statistical uncertainties reported in Table~\ref{tab:properties}. 

The ratio of \Civ/\Ciii\ emission allows a measurement of \Cppp/\Cpp\ abundance which is useful for assessing ionization correction factors. 
We assume that the \Civ\ emission is nebular in origin, noting that interstellar absorption or stellar emission can result in under- or over-estimates of the \Cppp\ abundance, respectively. Ultimately our reported results allow for up to a conservative factor of 2 uncertainty in \Cppp.
We follow the same approach as for \Cpp/\Opp, assuming the same temperature in the \Cppp\ zone and a relative reddening correction of $1.03\pm0.10$ (i.e., \Civ\ attenuated by 1.03$\times$ more than \OiiiUV\ and \Ciii). The abundance from \Civ~$\lambda$1549 flux is half that from $\lambda$1551, although consistent within 1.3$\sigma$. 
We view the $\lambda$1551 line as more reliable since it is less susceptible to absorption by interstellar and outflowing gas, and adopt this as our reference for the abundance. 
The resulting \Cppp/\Cpp\ abundance is given in Table~\ref{tab:properties} with uncertainties propagated as above. 


Our best measurement of metal ion abundance relative to hydrogen is \Opp/\Hp, which we obtain from the ratio of \Oiii~$\lambda$5007/\Hb\ using the \Te\ method. We report the value as \OHpp\ in Table~\ref{tab:properties}. The uncertainty is dominated by \Te.

\subsubsection{Ionization correction factor and total gas-phase abundances}

Having established the ionic abundances, we now turn to the total gas-phase abundances of C/O and O/H. This requires an ionization correction factor (ICF) for unobserved ions. 
The ionization correction from \Opp/\Hp\ to O/H is typically estimated using the reddening-corrected O32~=~\Oiii~$\lambda$5007/\Oii~$\lambda$3727 ratio alongside a relation to convert measured \Te(\Opp) to \Te(\Op). Since \Oii~$\lambda$3727 is not covered due to the wavelength range of the observations, we cannot directly measure O32. The \Oii~$\lambda$2471 line is covered but undetected, yielding a 3$\sigma$ upper limit of 12+log(O$^+$/H$^+)<7.35$. This limit suggests that more than half of O is in \Opp, as expected based on the relatively high \Cppp/\Cpp\ ratio. 
This oxygen ICF (i.e., \Opp/\Op~$>1$) in turn suggests an ICF from \Cpp/\Opp\ to total C/O of $\gtrsim0.75\times$ \citep[e.g.,][]{berg2019}. As this result is relatively unconstraining, we also consider indirect estimates for \Op.
It has been shown that O32 is strongly correlated with the rest-frame equivalent width of \Oiii~$\lambda$5007 (EW(5007)) over 2.5 orders of magnitude for star-forming galaxies at $z\sim2-3$ \citep{tang2019,sanders2020}. Using the spectroscopically measured EW(5007)~$=480\pm30$~\AA, the O32--EW(5007) relation of \citet{sanders2020} implies O32~$=3.6$. However, it is unclear whether this relation evolves between $z\sim2$ and $z>6$. The handful of published $z>6$ sources with O32 and EW(5007) measurements \citep[e.g.,][]{curti2023,trussler2022,schaerer2022} lie $0.1-0.3$~dex above the \citet{sanders2020} relation, suggesting O32~=~3.6 is a lower limit. The target of this analysis has \Oiii$\lambda$5007/H$\beta$~=~5.4, within the range measured by \citet{curti2023} for 3 galaxies at $z=7.5-8.5$ (\Oiii$\lambda$5007/H$\beta$~=~[3.08, 8.29, 7.11]). These galaxies have O32~=~[9.32, 8.94, 13.65]\footnote{We calculated O32 based on the observed line fluxes from \citet{curti2023} using the \cite{ccm89} dust curve assumed in this work. 
}, suggesting that O32 of our target falls in a similar range.

To estimate \Op/\Hp, we conservatively assume a uniform distribution of O32~$=3.0$--10 and adopt the median value of O32~$=6.5$ as our fiducial estimate. We further assume that \Te(\Op)~=~\Te(\Opp), though the resulting total O/H changes by $<0.05$~dex if we instead use the conversion of \citet{campbell1986} or \citet{pilyugin2009}. Under these assumptions, we estimate 12+log(O$^+$/H$^+)=6.34^{+0.33}_{-0.24}$. We then calculate the total O abundance under the common assumption O/H~$=\mathrm{\frac{O^{++} + O^{+}}{H^{+}}}$, finding 12+log(O/H$)=7.39^{+0.23}_{-0.20}$. The contribution from \Oppp\ is likely negligible in this case (and in all but the most extreme high-ionization sources). \citet{berg2018} used photoionization modeling to estimate an \Oppp\ fraction $\approx$0.05 for a high-ionization $z\sim2$ galaxy with \Cppp/\Cpp=0.86 (cf. $0.5\pm0.2$ in this work). The non-detection of \Heii\ also implies little \Oppp. Here, a $\le$5\% correction for \Oppp\ is significantly smaller than other sources of uncertainty.

Our target shows a significant contribution of both \Cpp\ and \Cppp, which provides useful information on the likely abundance of other ionic species of C. Photoionization models which reproduce this value of \Cppp/\Cpp\ \citep[e.g.,][]{berg2019} require high ionization parameters (log($U)\sim-1.5$) and low metallicity ($\lesssim$0.1--0.2$\times$ solar). These models indicate small contributions from other carbon ions ($\lesssim$10\% from \Cp). Oxygen ions in such cases are dominated by \Opp, with $\lesssim$10\% in the singly and triply ionized states consistent with the assumptions for O/H above. 
Furthermore, in the photoionization models of \citet{berg2019}, \Cp/C is nearly equal to \Op/O across the full range of grid points. 
Therefore we expect that ionic $\mathrm{\frac{C^{3+} + C^{++}}{O^{++}}}$ is approximately equal to total C/O abundance, with this approximation likely accurate to significantly better than 10\% for the case where neither \Cp\ nor \Op\ are observed.

We estimate that the sources of uncertainty in converting from ionic to total C/O abundance are of order 0.1~dex. The unseen states of C and O are likely of order $\sim$10\% as discussed above. The \Cppp\ abundance may be somewhat underestimated due to scattering and \Civ\ interstellar absorption (Section~\ref{sec:spectroscopy}), although underlying stellar wind emission could instead cause the \Cppp\ abundance to be overestimated. A factor of 2 change in \Cppp\ abundance corresponds to only 0.10~dex difference in the total $\mathrm{\frac{C^{3+} + C^{++}}{O^{++}}}$, which we view as a conservative limit. 
We therefore report the total C/O abundance in Table~\ref{tab:properties} as equal to the $\mathrm{\frac{C^{3+} + C^{++}}{O^{++}}}$ with an additional systematic uncertainty term.
We note that this value corresponds to an ICF from \Cpp/\Opp\ to total C/O of $0.17$ dex (or a factor 1.5$\times$), mainly driven by the \Cppp\ ion which we measure directly. 
This ICF(\Cpp/\Opp) value is consistent with the ICF function of \cite{amayo2021} for models with \Opp/O~$\approx0.90-0.95$ (cf. \Opp/O=$0.91\pm0.03$ based on O32 assumptions).
Allowing for a different relative \Te(\Cpp) by up to 2400~K (as discussed above) corresponds to $\pm0.15$~dex in \Cpp\ and $\pm0.11$~dex in C/O.
We sum this in quadrature with the ICF uncertainty and report the total systematic uncertainty as 0.15~dex in C/O abundance (Table~\ref{tab:properties}).

In summary, we have assessed various factors which affect the derived abundance patterns. Ultimately the C/O abundance is based primarily on the well-measured \Ciii/\OiiiUV\ flux ratio (which to rough approximation scales linearly with C/O), combined with an ICF based on \Civ/\Ciii.

\subsection{Stellar mass, age, and star formation history from SED fitting}
\label{sec:SED}

\begin{figure*}
\center
 \includegraphics[height=1.7in]{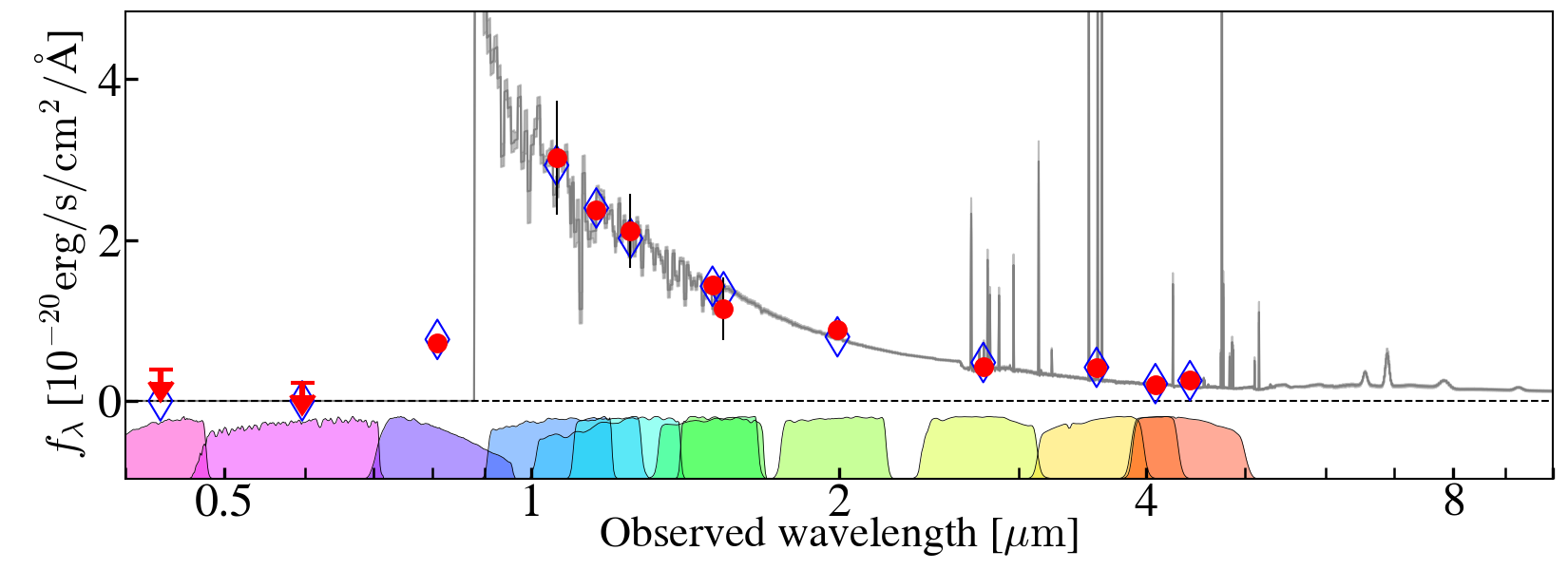}
 \includegraphics[height=2.0in]{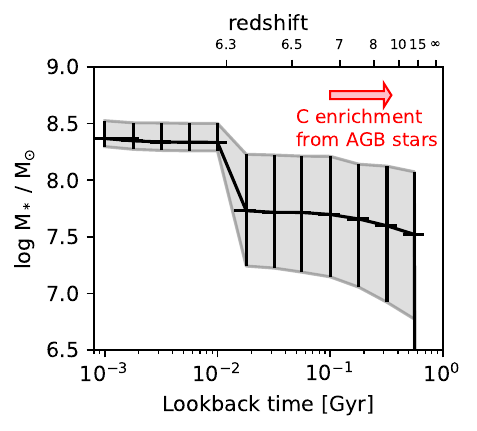}
 \caption{Results of SED fitting with a non-parametric star formation history \citep[for details see][]{morishita2019}.
 {\it Left:} The SED fit accurately reproduces the \hst\ and \jwst\ photometry (red points; open diamonds and grey spectrum show the best-fit model results), including strong emission line contributions in the F356W and F444W filters. The filter bandpasses are shown with colored shading below. 
 {\it Right:} Total stellar mass formed as a function of lookback time, based on the SED fit. The mass shown here is corrected for lensing magnification $\mu = 2.59$ (equivalent to 0.4 dex; we do not propagate the $\sim$1\% formal uncertainty in $\mu$ here). The $\gtrsim100$~Myr timescale for increased C/O from AGB stars is indicated with a red arrow, corresponding to stars formed at $z\gtrsim7$ in this case. 
 C/O abundance of this galaxy indicates enrichment dominated by core-collapse supernovae, with a majority of the stars formed within $\lesssim100$~Myr of the observed epoch, consistent with the best-fit mass-weighted age from SED fitting.}
 \label{fig:SED}
\end{figure*}

We fit the spectral energy distribution (SED) using the \jwst\ and \hst\ photometry (Table~\ref{tab:fluxes}; Figure~\ref{fig:SED}) following the methods of \cite{santini2022}, which demonstrated powerful constraints on the stellar mass of high redshift ($z\gtrsim7$) galaxies. We assume the \cite{bc03} stellar templates, a delayed-$\tau$ star formation history (SFH), \cite{chabrier2003} initial mass function (IMF), and \cite{calzetti2000} extinction law. 
This extinction law has the same $R_V$ as that adopted for our spectroscopic analysis, while we further allow for a range of UV attenuation curves (e.g., 2175~\AA\ bump strengths) in determining the C/O abundance in Section~\ref{sec:abundances}.
The best-fit stellar mass M$_*$, star formation rate (SFR), and stellar age are given in Table~\ref{tab:properties}. M$_*$ and SFR are corrected for lensing magnification. 

We refer readers to \cite{santini2022}, \cite{dressler2022}, \cite{whitler2023}, and references therein for further discussion of the SFH at high redshifts and the uncertainties associated with SED fitting. A main limitation at high redshifts is the availability of long-wavelength photometry, with \jwst/NIRCam providing reasonable sampling redward of the Balmer and 4000~\AA\ breaks up to $z\lesssim7.5$. The SED of our target is sampled with three filters redward of these breaks (Figure~\ref{fig:SED}), including F410M which is relatively free of strong emission lines and thus anchors the continuum flux density at rest-frame $\sim$5700~\AA.

The most relevant stellar population parameter for our abundance analysis is the stellar age. 
The SED fit with delayed-$\tau$ SFH favors a young $\simeq$130~Myr age, though the 1$\sigma$ confidence interval extends up to 500~Myr (corresponding to $z\simeq12$). 
To assess possible systematic uncertainties and better constrain the age, we performed an independent analysis of the same photometry with a non-parametric star formation history using the \textsc{gsf}\footnote{https://github.com/mtakahiro/gsf}
\citep{morishita2019} and {\it SEDz*} \citep{dressler2022} software packages. In brief, \textsc{gsf} fits the observed photometry with a linear combination of stellar population templates of different age bins, generated with the stellar population synthesis code {\tt fsps} \citep[][]{conroy09}, where each bin represents a short ($\sim$30 Myr) burst of constant SFR. 
{\it SEDz*} uses a combination of 10 Myr bursts and constant star formation templates, and is specifically designed to obtain SFHs for $z>5$ galaxies, taking advantage of the fact that their SEDs are largely dominated by class A stars.
Figure~\ref{fig:SED} shows the best-fit SED and star formation history (stellar mass formed per time bin) from \textsc{gsf}. 
The resulting mass-weighted age of $100^{+130}_{-60}$~Myr is fully consistent with that following \cite{santini2022}. The de-magnified $\log\,\mathrm{M}_* /\Msun = 8.37^{+0.16}_{-0.07}$
also agrees within the uncertainties. 
Likewise, {\it SEDz*} fitting results suggest that the majority of stellar mass formed within the preceding $\lesssim150$~Myr. 
We consider these non-parametric ages to be more reliable. We note that the best-fit stellar population ages are somewhat lower than the predicted average at $z\simeq6$ from \cite{mason2015}, which could be a result of selection bias for young age arising from the requirement of rest-UV emission line detections. 
Overall the three separate photometric analyses give a consistent picture but with relatively broad allowed ages, from a few tens to hundreds of Myr. When considering only the photometry, it is thus unclear whether the majority of stars seen in this galaxy formed at $z<7$ (as indicated by the best-fit ages) or at $z>8$--10 (allowed within the 1$\sigma$ bounds). The C/O chemical enrichment information from rest-UV spectroscopy therefore provides a powerful complementary constraint on the star formation history.

\section{Discussion and Conclusions}
\label{sec:conclusions}

\begin{figure*}
\center
 \includegraphics[width=0.7\textwidth]{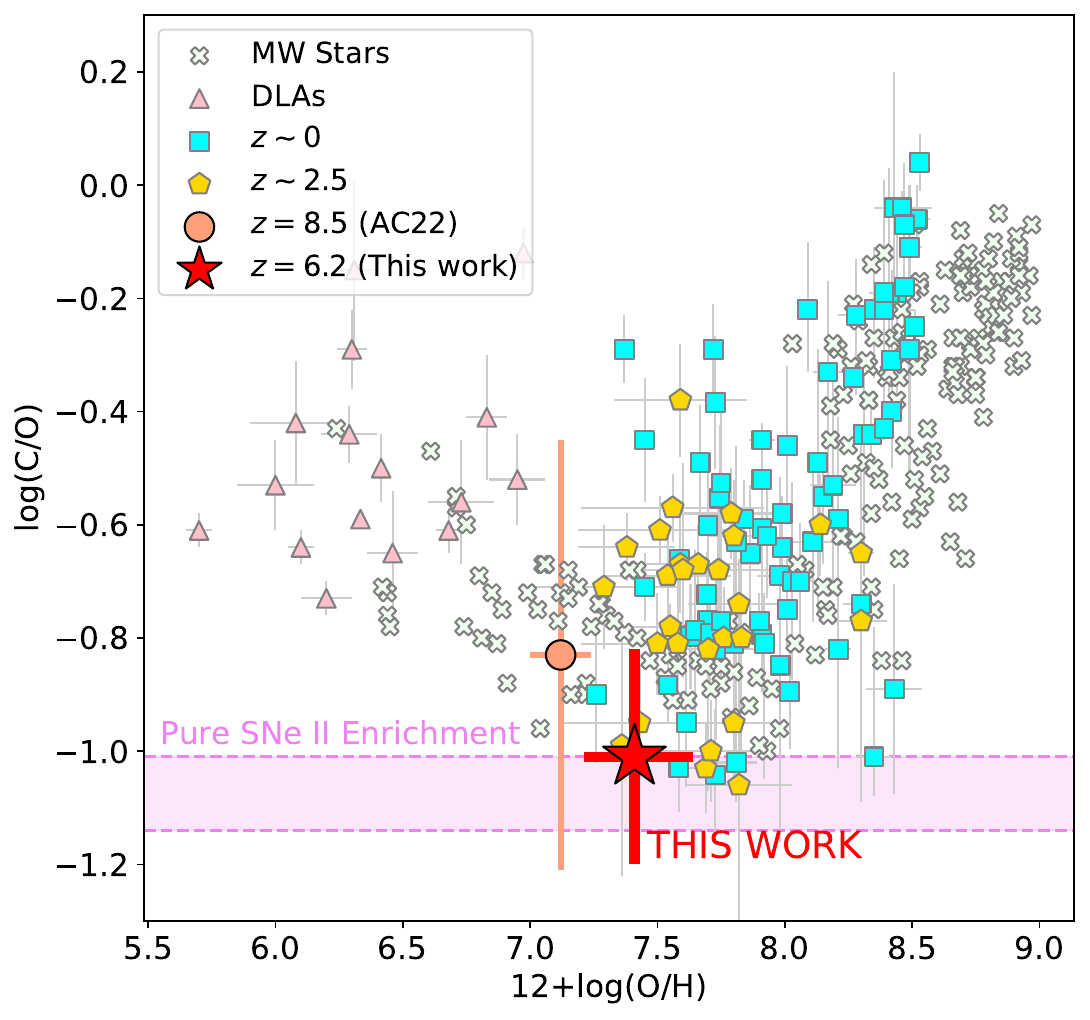}
 \caption{ The $\log (\mathrm{C/O)}$ and $12 + \log (\mathrm{O/H})$ values of our target $z=6.23$ galaxy (red star) compared to other objects compiled from literature: Milky Way halo and disk stars \citep[light-green crosses;][]{gustafsson99, akerman04, fabbian09, nissen14}; 
 damped Ly$\alpha$ absorbers \citep[DLAs, pink triangles;][]{cooke17}; local dwarf galaxies \citep{berg16, berg2019, penaguerrero17, senchyna17} and $z=0$ \Hii\ regions \citep{tsamis2003,garciarojas2004,garciarojas2005,garciarojas2006,garciarojas2007,esteban2004,esteban2009,esteban2014,esteban2017,peimbert2005,lopezsanchez2007,tsc16,tsc17}
 ($z \sim 0$; cyan squares); high-redshift galaxies near cosmic noon ($z \simeq 1.5$--3.5) \citep[orange pentagons;][]{fosbury03, erb10, christensen12, bayliss14, james14, stark14, steidel16, vanzella16, amorin17, berg18, mainali20, rigby21, matthee21, iani22}; and a galaxy at $z = 8.5$ from \citet{arellano22} (AC22: orange circle). The $\log (\mathrm{C/O})$ ratio from pure core-collapse SNe enrichment is marked with violet shading. This C/O range was calculated using the $Z_* = 0.05 Z_\odot$ and $0.2 Z_\odot$ values from the \citet{nomoto2013} yield tables assuming a \citet{salpeter1955} IMF. The C/O abundance ratio of this galaxy is at the lower envelope of other low-metallicity sources, consistent with pure SNe II enrichment and implying a young stellar population without significant enrichment from AGB stars.}  
 \label{fig:CO_OH}
 \footnotesize \textit{Note}--- For the $z=0$ \Hii\ regions, C/O was measured from recombination lines (RL). We plot O/H derived using the collisionally-excited line (CEL) \Te-method to match the O/H scale of the $z\sim0$ dwarf galaxy and $z\gtrsim2$ samples. We assume that the abundance discrepancy factor is the same for C and O \citep[e.g.,][]{tsc17}, such that C/O derived from RLs and CELs can be fairly compared.
\end{figure*}

We find a low gas-phase abundance ratio $\log{\mathrm{(C/O)}} = -1.01 \pm 0.12~\mathrm{(stat)} \pm 0.15~\mathrm{(sys)}$ derived primarily from rest-frame UV emission lines. This corresponds to [C/O]~$=-0.78$ relative to the solar scale from \cite{asplund21}. The overall chemical enrichment of O/H places it near current estimates of the mass-metallicity relation at $z\gtrsim6$ \citep[e.g.,][]{langeroodi2022,jones2020,ma2016}, although we caution that this relation is not yet well established at such high redshifts. 
Figure~\ref{fig:CO_OH} compares our C/O measurement\footnote{For display purposes we add the statistical and systematic uncertainty in quadrature, giving $\log{\mathrm{(C/O)}} = -1.01 \pm 0.19$.} at $z=6.23$ with other star forming galaxies and \Hii\ regions at $z\lesssim3$, the recent $z\approx8.5$ measurement from \cite{arellano22}, Milky Way stars, and damped \Lya\ systems measured from quasar spectra. 
Our measurement is in the lower envelope of known values from previous work, making this galaxy one of the lowest C/O systems known, and comparable to similarly metal-poor galaxies at lower redshifts.

A main point of interest is whether the galaxy is old enough to have undergone significant enrichment from intermediate-mass stars in their AGB phase, as opposed to being dominated by core-collapse supernova yields. We calculate the C/O abundance ratio expected from pure core-collapse SNe enrichment using values from \citet{nomoto2013} integrated over the massive end of the IMF, assuming yields for stellar metallicities $Z_* = 0.05$--$0.2~Z_\odot$ (shaded region in Figure~\ref{fig:CO_OH}). 
This $Z_*$ range corresponds to adjacent values in the \cite{nomoto2013} yield tables which straddle the metallicity of our target galaxy. Theoretical yield estimates vary with the initial metallicity which affects, e.g., the stellar evolution, pre-supernova mass loss, and rotational velocity. In this case higher initial metallicity results in larger predicted C/O yields.
The core-collapse C/O ratio lies at the extreme lower envelope of local and $z\sim2$ sources, such that the vast majority of known sources require additional enrichment from AGB stars. 

Our C/O measurement agrees with the SNe yield predictions (Figure~\ref{fig:CO_OH}), indicating minimal carbon enrichment from processes other than massive star supernovae. This implies a stellar population age $\lesssim$150~Myr. In older systems, we would expect to see enhanced carbon abundance due to significant enrichment from AGB stars.
Including the full yields from AGB stars down to 1~$\Msun$ \citep{nomoto2006} increases the C/O ratio by 0.5--0.6 dex compared to SNe alone, i.e., to $\log(\mathrm{C/O}) = -0.4$ to $-0.5$ for the same metallicity range shown in Figure~\ref{fig:CO_OH}. Considering only the yields from stars with main sequence lifetimes shorter than the age of the universe at $z=6.3$ (i.e., $\mathrm{M_*} >2.5$--$3~\Msun$) results in a $\sim$0.25 dex increase to $\log(\mathrm{C/O}) \approx -0.7$. While a 0.25 dex enhancement is compatible with our measurement, this assumes a closed-box chemical evolution with no inflow or outflow. For the closed-box case we can use the oxygen yields ($y_O \approx 0.038$ for the metallicity of our target and our adopted IMF; \citealt{vincenzo2016}) to infer a gas fraction $\mu = \frac{\mathrm{M_{gas}}}{\mathrm{M_{gas} + M_{*}}} = 0.89\pm0.05$. Given the measured stellar mass (Table~\ref{tab:properties}), the resulting total baryonic mass $M \simeq 2.2\times10^{9}~\Msun$ is several times larger than the dynamical mass estimated in Section~\ref{sec:spectroscopy}. This in turn suggests that a majority of the oxygen produced in SNe is ejected via outflows, which can increase C/O relative to the closed-box yields in the case where AGB enrichment has occurred. Incidentally the redshifted \Civ\ emission (Figure~\ref{fig:line_velocities}; Section~\ref{sec:spectroscopy}) is indicatative of a metal-enriched outflow in our target.

Chemical evolution models clearly indicate that the pure core-collapse ``floor'' C/O value can only be reached when there is no preferential removal of O relative to C from SNe-driven outflows, which is likely not the case when long-timescale AGB enrichment is present \citep[e.g.,][]{yin2011,berg2019}. These models suggest $\log{\mathrm{(C/O)}} \gtrsim -0.5$ following enrichment from AGB stars when modest outflows are included (e.g., when $\gtrsim$30\% of oxygen is ejected). Our results do not support such high C/O values.
This conclusion of little AGB contribution -- based on chemical abundance analysis -- is consistent with the results of SED fitting which likewise indicate a young age, although with large uncertainty ($100^{+130}_{-60}$~Myr; Section~\ref{sec:SED}, Figure~\ref{fig:SED}).
The $z\approx8.5$ source with log(C/O$)=-0.83\pm0.38$ from \citet{arellano22} is also inferred to have a very young age \citep[$<10$~Myr;][]{carnall23}, and is consistent with the pure core-collapse scenario, though the abundance constraint for this object is not robust due to the larger uncertainty.

Overall our results indicate a picture of rapid buildup of stellar mass in a galaxy seen only $\simeq$900~Myr after the big bang, with the majority of stellar mass assembled since $z\lesssim8$ (i.e., within 150~Myr) and likely even more recently. The current SFR and stellar mass suggest a mass doubling timescale of only $\lesssim$50~Myr, indicating a rising star formation history based on SED fitting (Figure~\ref{fig:SED}). 
It is therefore unlikely that this galaxy would have contributed significantly to cosmic reionization at $z\gtrsim8$, as our C/O abundance analysis indicates little star formation ($\lesssim 10^8~\Msun$ total mass) before this time. 

As our results represent the most robust C/O abundance to date in such a high-redshift galaxy, we reflect briefly on lessons learned and prospects for future study. Sensitive rest-UV spectroscopy is essential for this result; the relatively small uncertainty is thanks to clear detection of the UV \OiiiUV, \Ciii, and \Civ\ emission doublets. Notably, we find that \Civ\ is important for assessing the ionization correction factor in this case (where we find \Cppp\ contributes at the level of 0.17 dex), whereas it is often not significant in lower-redshift samples. 
Care should be taken to assess possible interstellar absorption and stellar contributions to the \Civ\ profile.
One of the larger sources of uncertainty is the relative \Te\ associated with emission lines of different ions, as discussed in Section~\ref{sec:abundances}, which warrants further examination to reach precision better than $\simeq0.1$ dex.
The precision of our measurement could also be improved with the addition of \Oii~$\lambda\lambda$3726,3729, not covered in our observations, that would yield improved estimates of the ICF. 
The \Oii\ doublet would also provide a better measurement of electron density \Ne.
Measurements of rest-optical \Oiii~$\lambda$4363 (falling in the chip gap in our G395H observations) would provide a better constraint on \Te\ relative to our value based on rest-UV \OiiiUV~$\lambda$1666, for which the error budget is dominated by uncertainty on the reddening correction. 
The main consequence of these missing emission line diagnostics is that total O/H (and C/H) abundances have larger uncertainty; the effect on derived C/O abundance is relatively minor. 
Nonetheless it is fully within the capabilities of \jwst/NIRSpec to provide these additional measurements with an appropriate mask and filter configuration. Our result thus represents only a lower limit to the C/O precision that can be achieved at $z>6$ with NIRSpec spectroscopy.
There is also room for improvement in chemical evolution modeling, which has largely focused on abundance patterns at lower redshifts \citep[e.g.][]{yin2011,berg2019,kobayashi2020}. The results of this work and other high-redshift abundance analyses \citep[e.g.,][]{arellano22,cameron2023} motivate exploring multi-element models which are specifically tailored to the rapid formation histories expected in the first billion years of the Universe.

This work demonstrates the value of gas-phase C/O abundance for characterizing star formation histories of galaxies in the epoch of reionization, and the feasibility of reaching good precision with \jwst\ data. A larger sample of $z>6$ targets with C/O measurements will be valuable to characterize the typical SFHs and to compare with complementary results from photometry and SED fitting. 
For example, \cite{laporte2022} find stellar population ages $>150$~Myr in 2 galaxies at $z>8$ within a sample of 6 based on \jwst\ photometry, improving upon earlier Spitzer-based results \citep[e.g.,][]{rb20}. We would expect these older galaxies to exhibit higher C/O. 
A positive correlation between C/O and photometrically derived ages \citep[e.g.,][]{dressler2022} would bolster confidence in both methods. 
If instead rapid formation histories and ages $\lesssim150$~Myr are a uniform feature of $z>6$ galaxies, then we expect the population to display lower average C/O and smaller intrinsic scatter in C/O at fixed O/H relative to samples at $z\sim0$ and $z\sim2-3$ (Figure~\ref{fig:CO_OH}). 
Our results motivate the assembly of a larger sample of reionization-era targets with robust rest-UV \Ciii\ and \OiiiUV\ measurements to constrain the timescale of galaxy assembly in the early universe.


\acknowledgments
We thank Michael Topping for allowing a comparison of the line fluxes derived in this work to those measured from an independent data reduction.
We are grateful to the referee for providing a constructive review which improved this manuscript.
This work is based on observations made with the NASA/ESA/CSA James Webb Space Telescope. This work also uses observations made with the NASA/ESA Hubble Space Telescope. 
The data were obtained from the Mikulski Archive for Space Telescopes at the Space Telescope Science Institute, which is operated by the Association of Universities for Research in Astronomy, Inc., under NASA contract NAS 5-03127 for JWST. The NIRSpec and NIRCam observations are associated with programs JWST-ERS-1324 and JWST-GO-2561, respectively. 
We acknowledge financial support for program GLASS-JWST ERS-1324 provided by NASA through grant JWST-ERS-1324 from the Space Telescope Science Institute. 
Support for this work was provided by NASA through the NASA Hubble Fellowship grant HST-HF2-51469.001-A awarded by the Space Telescope Science Institute, which is operated by the Association of Universities for Research in Astronomy, Incorporated, under NASA contract NAS5-26555. We acknowledge support from the INAF Large Grant 2022 “Extragalactic Surveys with JWST”  (PI Pentericci).
XW is supported by CAS Project for Young Scientists in Basic Research, Grant No. YSBR-062. MB acknowledges support from the Slovenian national research agency ARRS through grant N1-0238.
KG and TN acknowledge support from Australian Research Council Laureate Fellowship FL180100060. 
The Cosmic Dawn Center is funded by the Danish National Research Foundation (DNRF) under grant \#140.  Cloud-based data processing and file storage for this work is provided by the AWS Cloud Credits for Research program.




\bibliography{nirspec_CtoO}{}
\bibliographystyle{aasjournal}

\end{document}